\documentstyle[11pt]{article}
\newenvironment{eqnarraz}%
   {\setlength{\arraycolsep}{0.15em} \begin{eqnarray}}%
   {\end{eqnarray}}
\begin{document}
\vspace{20mm}
\begin{center}
{\LARGE \bf Lie Symmetries of Einstein's Vacuum Equations\\[2mm]
in $N$ Dimensions}\\[5mm]
Louis Marchildon\\
D\'{e}partement de physique, Universit\'{e} du Qu\'{e}bec,\\
Trois-Rivi\`{e}res, Qu\'{e}bec, Canada G9A 5H7\\
FAX: (819) 376-5164; e-mail: marchild@uqtr.uquebec.ca\\[5mm]
\end{center}
\begin{abstract}
We investigate Lie symmetries of Einstein's
vacuum equations in $N$ dimensions, with a cosmological term.
For this purpose, we first write down the second prolongation of
the symmetry generating vector fields, and compute its action on
Einstein's equations.  Instead of setting to zero the
coefficients of all independent partial derivatives (which
involves a very complicated substitution of Einstein's
equations), we set to zero the coefficients of derivatives that
do not appear in Einstein's equations.  This considerably
constrains the coefficients of symmetry generating vector fields.
Using the Lie algebra property of generators of symmetries and
the fact that general coordinate transformations are symmetries
of Einstein's equations, we are then able to obtain all the Lie
symmetries. The method we have used can likely be applied to
other types of equations.  PACS: 02.20.+b
\end{abstract}
\section{Introduction}
Consider a nondegenerate system of {\it n}-th order
nonlinear partial differential equations for a number of
independent variables $x$ and dependent variables $g$:
\begin{equation}
\Delta _{\nu} (x,g,\partial g, \ldots ,\partial ^{(n)} g) = 0.
\label{delta}
\end{equation}
Let $v$ be a linear combination of first-order partial
derivatives with respect to the $x$ and $g$, with coefficients
depending on the $x$ and $g$.  Then $v$ will generate a Lie
symmetry of Eq.~(\ref{delta}) if and only if the following
holds~\cite{olver}:
\begin{equation}
\left[ \mbox{pr} ^{(n)} v \right] \Delta _{\nu} = 0
\;\; \mbox{whenever} \;\; \Delta _{\nu} = 0 ,
\label{prdelta}
\end{equation}
where $\mbox{pr} ^{(n)} v$ is the so-called {\it n}-th
prolongation of $v$.  Eq.~(\ref{prdelta}) constitutes a
system of
linear equations for the coefficients of partial derivatives
making up the operator $v$.

To compute Eq.~(\ref{prdelta}) explicitly, the main problem
consists in eliminating nonindependent partial derivatives
through substitution of $\Delta _{\nu} =  0$.  This can be
complicated, as illustrated by the case of the Yang-Mills
equations examined elsewhere~\cite{marchildon}.

In this paper, we shall investigate Lie symmetries of Einstein's
vacuum equations in {\it N} dimensions, including a cosmological
term.  Substitution of Eq.~(\ref{delta}) is much more
complicated
here than in the Yang-Mills case.  We will show, however, that
the substitution can be bypassed by using the Lie algebra
property of symmetry generators and knowledge of some of the
symmetries.  This technique can likely be used in other systems
of nonlinear partial differential equations.

In Section~2, we write down Einstein's vacuum equations in {\it
N} dimensions (Einstein's equations, for short), and recall some
of their properties.  In Section~3, we compute the action of
the second prolongation of $v$ on Einstein's equations.
Coefficients of partial derivatives not appearing in Einstein's
equations must vanish identically, and this is effected in
Section~4\@.  There result constraints on symmetry generators
which, however, are not enough to determine the generators
completely.  In Section~5, we use the fact that general
coordinate transformations are symmetries of Einstein's
equations, together with the Lie algebra property of generators
of symmetries, to show that the complete set of Lie symmetries of
Einstein's equations coincides with general
coordinate transformations and, when the cosmological term
vanishes, uniform rescalings of the metric.

Lie symmetries~\cite{ibragimov}
and generalized symmetries~\cite{torre,torrea}
of the Einstein vacuum equations in 4 dimensions,
without a cosmological term, were investigated
before, with results in agreement with ours.
\section{Einstein's vacuum equations in $N$ dimensions}
Einstein's vacuum equations in $N$ dimensions can be written
as~\cite{landau,weinberg,brown}
\begin{equation}
R_{\alpha \beta} - \lambda g_{\alpha \beta} = 0 .
\label{eve}
\end{equation}
Here $\lambda$ is a constant, and $\lambda g_{\alpha \beta}$  is
the cosmological term.  $R_{\alpha \beta}$, the Ricci tensor, is
given by
{\begin{eqnarraz}
R_{\alpha \beta} &=& \frac{1}{2} g^{\gamma \delta} \left\{
- \partial _{\gamma} \partial _{\delta} g_{\alpha \beta}
- \partial _{\alpha} \partial _{\beta} g_{\gamma \delta}
+ \partial _{\beta} \partial _{\delta} g_{\alpha \gamma}
+ \partial _{\alpha} \partial _{\gamma} g_{\delta \beta} \right\}
\nonumber \\
& & \quad \mbox{} + g^{\gamma \delta}
g^{\tau \rho} \left\{ \Gamma _{\tau
\gamma \alpha} \Gamma _{\rho \delta \beta} - \Gamma _{\tau \gamma
\delta} \Gamma _{\rho \alpha \beta} \right\}.
\label{ricci}
\end{eqnarraz}}%
The symbol $\partial _{\gamma}$ represents a partial derivative
with respect to the independent variable $x^{\gamma}$
$(\gamma = 1,
\ldots , N)$.  The $g_{\nu \lambda} = g_{\lambda \nu}$ are
dependent variables.  The $g^{\mu \nu}$ are defined so that
\begin{equation}
g^{\mu \nu} g_{\nu \lambda} = \delta _{\mu} ^{\lambda} ,
\label{kronecker}
\end{equation}
where $\delta _{\mu} ^{\lambda}$ is the Kronecker delta.  The
Christoffel symbols $\Gamma _{\tau \gamma \alpha}$ are given by
\begin{equation}
\Gamma _{\tau \gamma \alpha} = \frac{1}{2} \left\{ \partial
_{\alpha} g_{\tau \gamma}  + \partial _{\gamma} g_{\tau \alpha} -
\partial _{\tau} g_{\gamma \alpha} \right\}. \label{christoffel}
\end{equation}
Note that we have
\begin{equation}
\partial _{\alpha} g _{\tau \gamma} = \Gamma _{\tau \gamma
\alpha} + \Gamma _{\gamma \tau  \alpha} . \label{christoffel1}
\end{equation}

There are $N(N+1)/2$ variables $g_{\nu \lambda}$.  Varying them
independently, we get from Eq.~(\ref{kronecker})
\begin{equation}
\delta g^{\mu \nu} = - g^{\mu \kappa} (\delta g_{\kappa \lambda})
g^{\nu \lambda} ,
\end{equation}
whence
\begin{equation}
- \frac{\partial g^{\mu \nu}}{\partial g_{\kappa \lambda}} =
X^{\mu \nu \kappa \lambda} \equiv \left\{
  \begin{array}{cl}
  g^{\mu \kappa} g^{\nu \lambda} + g^{\mu \lambda} g^{\nu \kappa}
     & \mbox{ if } \kappa \neq \lambda , \\
  g^{\mu \kappa} g^{\nu \lambda} & \mbox{ if }
    \kappa = \lambda .
  \end{array} \right.
\label{x}
\end{equation}
In the symbol $X^{\mu \nu \kappa \lambda}$, indices can be
lowered, for instance
\begin{equation}
{X_{\mu \nu}}^ {\kappa \lambda} = \left\{
  \begin{array}{cl}
  \delta _{\mu}^{\kappa} \delta _{\nu}^{\lambda} +
  \delta _{\mu}^{\lambda} \delta _{\nu}^{\kappa}
  & \mbox{ if } \kappa \neq \lambda , \\
  \delta _{\mu}^{\kappa} \delta _{\nu}^{\lambda} & \mbox{ if }
  \kappa = \lambda .
  \end{array} \right.
\label{xl}
\end{equation}
Note that we have
\begin{equation}
\frac{\partial g_{\mu \nu}}{\partial g_{\kappa \lambda}} =
{X_{\mu \nu}}^ {\kappa \lambda}.
\label{xll}
\end{equation}

For later purposes, we now evaluate the partial derivatives of
the Ricci tensor with respect to the metric tensor and its
partial derivatives.  For this, the $X$ symbol is particularly
useful.  From Eqs.~(\ref{ricci}), (\ref{x}) and
(\ref{xll}), we
find that
\begin{equation}
 \frac{\partial R_{\alpha \beta}} {\partial (\partial
_{\kappa} \partial _{\lambda} g_{\mu \nu})} = \frac{1}{2}
g^{\gamma \delta}
\left\{ - {X_{\gamma \delta}} ^{\kappa \lambda} {X_{\alpha
\beta}} ^{\mu \nu}- {X_{\alpha \beta}} ^{\kappa \lambda}
{X_{\gamma \delta}} ^{\mu \nu} + {X_{\delta \beta}} ^{\kappa
\lambda} {X_{\gamma \alpha}} ^{\mu \nu} + {X_{\gamma \alpha}}
^{\kappa \lambda} {X_{\delta \beta}} ^{\mu \nu} \right\} ,
\label{rddg}
\end{equation}
{\begin{eqnarraz}
\frac{\partial R_{\alpha \beta}} {\partial (\partial
_{\kappa}
g_{\mu \nu})} &=& \frac{1}{2} g^{\gamma \delta} g^{\tau \rho}
\left\{ \left[ \delta _{\alpha} ^{\kappa} {X_{\tau \gamma}} ^{\mu
\nu} + \delta _{\gamma} ^{\kappa} {X_{\tau \alpha}} ^{\mu \nu} -
\delta _{\tau} ^{\kappa} {X_{\gamma \alpha}} ^{\mu \nu} \right]
\Gamma _{\rho \delta \beta} \right. \nonumber \\
& & \quad \mbox{} + \left[ \delta _{\beta} ^{\kappa} {X_{\rho
\delta}} ^{\mu \nu} +
\delta _{\delta} ^{\kappa} {X_{\rho \beta}} ^{\mu \nu} - \delta
_{\rho} ^{\kappa} {X_{\delta \beta}} ^{\mu \nu} \right] \Gamma
_{\tau \gamma \alpha} \nonumber \\
& & \quad \mbox{} - \left[ \delta _{\delta} ^{\kappa} {X_{\tau
\gamma}} ^{\mu \nu}
+ \delta _{\gamma} ^{\kappa} {X_{\tau \delta}} ^{\mu \nu} -
\delta _{\tau} ^{\kappa} {X_{\gamma \delta}} ^{\mu \nu} \right]
\Gamma _{\rho \alpha \beta} \nonumber \\
& & \quad \mbox{} - \left. \left[ \delta _{\beta} ^{\kappa}
{X_{\rho \alpha}} ^{\mu
\nu} + \delta _{\alpha} ^{\kappa} {X_{\rho \beta}} ^{\mu \nu} -
\delta _{\rho} ^{\kappa} {X_{\alpha \beta}} ^{\mu \nu} \right]
\Gamma _{\tau \gamma \delta} \right\} \label{rdg} ,
\end{eqnarraz}}%
{\begin{eqnarraz}
 \frac{\partial R_{\alpha \beta}} {\partial g_{\mu \nu}} &=&
\frac{1}{2} \left\{ \partial _{\gamma} \partial _{\delta} g
_{\alpha \beta} + \partial _{\alpha} \partial _{\beta} g_{\gamma
\delta} - \partial _{\delta} \partial _{\beta} g_{\gamma \alpha}
- \partial _{\gamma}  \partial _{\alpha}  g_{\delta \beta}
\right\} X^{\gamma \delta \mu \nu} \nonumber \\
& & \quad \mbox{}- \left\{
\Gamma _{\tau \gamma \alpha} \Gamma _{\rho
\delta \beta} - \Gamma _{\tau \gamma \delta}  \Gamma _{\rho
\alpha \beta} \right\} \left\{ g^{\gamma \delta} X^{\tau \rho \mu
\nu} + g^{\tau \rho} X^{\gamma \delta  \mu \nu} \right\}.
\label{rg}
\end{eqnarraz}}%

A word on notations.  The summation convention on repeated
indices has hitherto been used.  There will, however, be
instances where we will not want to use it.  Following
\cite{marchildon}, we will put
carets on indices wherever summation should not be carried out.
This means, for instance, that in an equation like
\begin{equation}
M_{\mu} ^{\mu} = N_{\hat{\alpha}} ^{\hat{\alpha}} ,
\end{equation}
summation is carried out over $\mu$ but not over $\hat{\alpha}$,
the latter index having a specific value.  Furthermore, when
dealing with symmetric matrices we will often need to restrict a
summation to distinct values of a pair of indices.  In that case,
parentheses will enclose the indices, for instance
\begin{equation}
A_{(\mu \nu)} B^{\mu \nu} \equiv \sum _{\mu \leq \nu} A_{\mu \nu}
B^{\mu \nu}.
\end{equation}
Note that we have, for any symmetric $A$
\begin{equation}
A_{(\mu \nu)} {X_{\gamma \delta}} ^{\mu \nu} = A_{\gamma \delta}.
\label{sumx}
\end{equation}

In closing this section, we should point out that not all
second-order partial derivatives of the metric tensor appear in
the Ricci tensor.  Indeed writing down the
second-order derivatives explicitly, one can see that for any
values of $\rho$ and $\hat{\sigma}$, no terms like $\partial
_{\rho} \partial _{\hat{\sigma}} g_{\hat{\sigma} \hat{\sigma}}$
or $\partial _{\hat{\sigma}} \partial _{\hat{\sigma}} g_{\rho
\hat{\sigma}}$ appear in Eq.~(\ref{ricci}).
\section{Prolongation of vector fields}
The generator of a Lie symmetry of Einstein's vacuum equations in
$N$ dimensions has the form
\begin{equation}
v = H^{\mu} \frac{\partial}{\partial x^{\mu}} + \Phi _{(\mu \nu)}
\frac{\partial}{\partial g_{\mu \nu}}. \label{v}
\end{equation}
Here $H^{\mu}$ and $\Phi _{\mu \nu}$  are functions of the
independent variables $x^{\lambda}$ and dependent variables
$g_{\alpha \beta}$.  The second summation on the right-hand side
is restricted so that dependent variables are not counted twice.
Nevertheless, it is useful to define $\Phi _{\mu \nu}$ for $\mu >
\nu$ also, by setting $\Phi _{\mu \nu} = \Phi _{\nu \mu}$.

The second prolongation of $v$ is given by~\cite{olver}
\begin{equation}
 \mbox{pr} ^{(2)} v = H^{\mu} \frac{\partial}{\partial
x^{\mu}} +
\Phi _{(\mu \nu)} \frac{\partial}{\partial g_{\mu \nu}} + \Phi
_{(\mu \nu) \kappa} \frac{\partial}{\partial (\partial _{\kappa}
g_{\mu \nu})} + \Phi _{(\mu \nu) (\kappa \lambda)}
\frac{\partial}{\partial (\partial _{\kappa} \partial _{\lambda}
g_{\mu \nu})} . \label{prv}
\end{equation}
Here also, summations are restricted so that identical objects
are not counted twice.  The $\Phi _{\mu \nu \kappa}$ and $\Phi
_{\mu \nu \kappa \lambda}$ are functions of the independent and
dependent variables that will soon be examined.  Again, it is
useful to extend the range of indices so that
\begin{equation}
\Phi _{\mu \nu \kappa} = \Phi _{\nu \mu \kappa} \;\;
\mbox{and} \;\; \Phi
_{\mu \nu \kappa \lambda} = \Phi _{\mu \nu \lambda \kappa} = \Phi
_{\nu \mu \kappa \lambda} . \label{indsym}
\end{equation}
We now apply the right-hand side of Eq.~(\ref{prv}) on the
left-hand side of Eq.~(\ref{eve}), and substitute
Eqs.~(\ref{xll}), (\ref{rddg}), (\ref{rdg}) and (\ref{rg}).
Making use of Eq.~(\ref{sumx}) and rearranging, we find that
all restrictions on summations disappear, and we obtain
\begin{eqnarray}
\lefteqn{ \left[ \mbox{pr}^{(2)} v \right] \left\{ R_{\alpha
\beta} - \lambda g_{\alpha \beta} \right\} = - \lambda
\Phi _{\alpha \beta} + \frac{1}{2} \Phi ^{\gamma \delta} \left\{
\partial _{\gamma} \partial _{\delta} g _{\alpha \beta} +
\partial _{\alpha} \partial _{\beta} g_{\gamma \delta} - \partial
_{\delta} \partial _{\beta} g_{\gamma \alpha} - \partial
_{\gamma}  \partial _{\alpha}  g_{\delta \beta} \right\}}
\nonumber \\
& & \mbox{} - \left\{
\Gamma _{\tau \gamma \alpha} \Gamma _{\rho \delta
\beta} - \Gamma _{\tau \gamma \delta}  \Gamma _{\rho \alpha
\beta} \right\} \left\{ g^{\gamma \delta} \Phi ^{\tau \rho} +
g^{\tau \rho} \Phi^{\gamma \delta} \right\} \nonumber  \\
& & \mbox{} + \frac{1}{2} g^{\gamma \delta} g^{\tau \rho}
\left\{ \left[ \Phi _{\tau \gamma \alpha}
+ \Phi _{\tau \alpha \gamma} - \Phi _{\gamma \alpha \tau} \right]
\Gamma _{\rho \delta \beta} + \left[ \Phi _{\rho \delta \beta} +
\Phi _{\rho \beta \delta} - \Phi _{\delta \beta \rho} \right]
\Gamma _{\tau \gamma \alpha} \right. \nonumber \\
& & \quad \mbox{} - \left. \left[
\Phi _{\tau \gamma \delta} + \Phi _{\tau \delta
\gamma}  - \Phi _{\gamma \delta \tau} \right] \Gamma _{\rho
\alpha \beta} - \left[ \Phi _{\rho \alpha \beta}
+ \Phi _{\rho \beta \alpha} - \Phi _{\alpha \beta \rho} \right]
\Gamma _{\tau \gamma \delta} \right\} \nonumber \\
& & \mbox{} + \frac{1}{2} g^{\gamma \delta}
\left\{ -  \Phi _{\alpha
\beta \gamma \delta} - \Phi _{\gamma \delta \alpha \beta} + \Phi
_{\gamma \alpha \delta \beta} + \Phi _{\delta \beta \gamma
\alpha} \right\} . \hspace{57mm} \label{prvx}
\end{eqnarray}

The functions $\Phi_{\tau \gamma \alpha}$ and $\Phi _{\alpha
\beta  \gamma \delta}$ are given by~\cite{olver}
\begin{equation}
\Phi _{\tau \gamma \alpha} = D_{\alpha} \left\{\Phi _{\tau
\gamma} - H^{\eta} \partial _{\eta} g_{\tau  \gamma} \right\} +
H^{\eta} \partial _{\alpha} \partial _{\eta} g_{\tau  \gamma}
\label{phi3}
\end{equation}
and
\begin{equation}
\Phi _{\alpha \beta \gamma \delta} = D_{\gamma} D_{\delta}
\left\{ \Phi _{\alpha \beta} - H^{\eta} \partial _{\eta}
g_{\alpha \beta} \right\} + H^{\eta} \partial _{\gamma} \partial
_{\delta} \partial _{\eta} g_{\alpha \beta} \label{phi4}.
\end{equation}
Here $D_{\alpha}$ is the total derivative operator, given by
\begin{equation}
 D_{\alpha} = \partial _{\alpha} + \partial _{\alpha} g_{(\mu
\nu)} \frac{\partial }{\partial g_{\mu \nu}} + \partial _{\alpha}
\partial _{\kappa} g_{(\mu \nu)} \frac{\partial }{\partial
(\partial _{\kappa} g_{\mu \nu})} + \partial _{\alpha} \partial
_{( \kappa} \partial _{\lambda )} g_{(\mu \nu)} \frac{\partial
}{\partial (\partial _{\kappa} \partial _{\lambda} g_{\mu \nu})}.
\label{totald}
\end{equation}
Substituting Eq.~(\ref{totald}) in (\ref{phi3}) and
(\ref{phi4}), we obtain
\begin{equation}
\Phi _{\tau \gamma \alpha} = \partial_{\alpha} \Phi _{\tau
\gamma} - [\partial _{\eta} g_{\tau  \gamma}] \partial _{\alpha}
H^{\eta} + \partial _{\alpha} g_{(\mu \nu)} \frac{\partial \Phi
_{\tau \gamma}}{\partial g_{\mu \nu}} - [\partial _{\alpha}
g_{(\mu \nu)}] \partial _{\eta}  g_{\tau \gamma} \frac{\partial H
^{\eta}}{\partial g_{\mu \nu}} \label{phi3x}
\end{equation}
and
{\begin{eqnarraz}
\Phi _{\alpha \beta \gamma  \delta} &=& \partial
_{\gamma} \partial
_{\delta} \Phi _{\alpha \beta} - [\partial _{\eta}  g_{\alpha
\beta}] \partial _{\gamma} \partial _{\delta} H^{\eta} + \partial
_{\delta} g_{(\mu \nu)} \partial _{\gamma} \left( \frac{\partial
\Phi _{\alpha \beta}}{\partial g_{\mu \nu}} \right) + \partial
_{\gamma} g_{(\mu \nu)} \partial _{\delta} \left( \frac{\partial
\Phi _{\alpha \beta}}{\partial g_{\mu \nu}} \right) \nonumber \\
& & \quad \mbox{}
- [\partial _{\gamma} g_{(\mu \nu)}] \partial _{\eta}
g_{\alpha \beta} \partial _{\delta} \left( \frac{\partial
H^{\eta}}{\partial g_{\mu \nu}} \right) - [\partial _{\delta}
g_{(\mu \nu)}] \partial _{\eta} g_{\alpha \beta} \partial
_{\gamma} \left( \frac{\partial H^{\eta} }{\partial g_{\mu \nu}}
\right) \nonumber \\
& & \quad \mbox{} + [\partial _{\gamma}
g_{(\mu \nu)}] \partial _{\delta}
g_{(\pi \sigma)} \frac{\partial ^2 \Phi _{\alpha \beta}
}{\partial g_{\mu \nu} \partial g_{\pi \sigma}} - [\partial
_{\gamma} g_{(\mu \nu)}] [\partial _{\delta}
g_{(\pi \sigma)}] \partial _{\eta} g_{\alpha \beta}
\frac{\partial ^2 H^{\eta} }{\partial g_{\mu \nu} \partial g_{\pi
\sigma}} \nonumber \\
& & \quad \mbox{} - [\partial _{\delta}
\partial _{\eta} g_{\alpha \beta}]
\partial _{\gamma}  H^{\eta} - [\partial _{\gamma} \partial
_{\eta} g_{\alpha \beta}] \partial _{\delta} H^{\eta} + \partial
_{\gamma} \partial _{\delta} g_{(\mu \nu)} \frac{\partial \Phi
_{\alpha \beta}}{ \partial g _{\mu \nu}} \nonumber \\
& & \quad  \mbox{} - [\partial _{\gamma}
g_{(\mu \nu)}] \partial _{\delta}
\partial _{\eta} g_{\alpha \beta} \frac{\partial H^{\eta}}{
\partial g_{\mu \nu}} - [\partial _{\delta} g_{(\mu \nu)}]
\partial _{\gamma} \partial _{\eta} g_{\alpha \beta}
\frac{\partial H^{\eta}}{ \partial g_{\mu \nu}} \nonumber \\
& & \quad \mbox{} - [\partial
_{\eta} g_{\alpha \beta}] \partial _{\delta} \partial _{\gamma}
g_{(\mu \nu)} \frac{\partial H^{\eta}}{  \partial g_{\mu \nu}}
. \label{phi4x}
\end{eqnarraz}}%
Note that Eqs.~(\ref{phi3x}) and (\ref{phi4x}) are
consistent
with (\ref{indsym}).

The action of the second prolongation of $v$ on the
left-hand side of Eq.~(\ref{eve})
can now be obtained by substituting (\ref{phi3x}) and
(\ref{phi4x}) into (\ref{prvx}).  The resulting equation
is very complicated but, fortunately, we will not have to write
it down all at once. In any case, the conditions under which it
vanishes, subject to Eq.~(\ref{eve}), must be found so as to
determine the Lie symmetries of (\ref{eve}).
\section{Determining equations}
In this section, we will consider in turn several combinations of
partial derivatives of $g_{\mu \nu}$ appearing in
Eq.~(\ref{prvx}).
\subsection*{$\partial g \partial \partial g$ terms}
There are three groups of $\partial g \partial \partial g$ terms
in Eq.~(\ref{phi4x}).  When they are substituted in
(\ref{prvx}), that makes altogether twelve groups of terms given
by
{\begin{eqnarraz}
 \partial g \partial \partial g & \rightarrow & \frac{1}{2}
g^{\gamma
\delta} \frac{\partial H^{\eta}}{\partial g_{(\mu \nu})} \left\{
\partial _{\gamma} g_{\mu \nu}  \partial _{\delta} \partial
_{\eta} g_{\alpha \beta} + \partial _{\delta} g_{\mu \nu}
\partial _{\gamma} \partial _{\eta} g_{\alpha \beta} + \partial
_{\eta} g_{\alpha \beta}  \partial _{\delta} \partial _{\gamma}
g_{\mu \nu} \right. \nonumber \\
& & \quad \mbox{} + \partial _{\alpha}
g_{\mu \nu}  \partial _{\beta}
\partial _{\eta} g_{\gamma \delta} + \partial _{\beta} g_{\mu
\nu} \partial _{\alpha} \partial _{\eta} g_{\gamma \delta} +
\partial _{\eta} g_{\gamma \delta}  \partial _{\alpha} \partial
_{\beta} g_{\mu \nu} \nonumber \\
& & \quad \mbox{} - \partial _{\delta}
g_{\mu \nu}  \partial _{\beta}
\partial _{\eta} g_{\alpha \gamma} - \partial _{\beta} g_{\mu
\nu} \partial _{\delta} \partial _{\eta} g_{\alpha \gamma} -
\partial _{\eta} g_{\alpha \gamma}  \partial _{\delta} \partial
_{\beta} g_{\mu \nu} \nonumber \\
& & \quad \mbox{}
- \left. \partial _{\gamma} g_{\mu \nu}  \partial
_{\alpha} \partial _{\eta} g_{\delta \beta} - \partial _{\alpha}
g_{\mu \nu}  \partial _{\gamma} \partial _{\eta} g_{\delta \beta}
- \partial _{\eta} g_{\delta \beta}  \partial _{\alpha} \partial
_{\gamma} g_{\mu \nu} \right\} . \label{dgddg1}
\end{eqnarraz}}%
Rearranging indices, one can show that Eq.~(\ref{dgddg1})
becomes
{\begin{eqnarraz}
\partial g \partial \partial g & \rightarrow & \frac{1}{2}
\partial
_{\gamma} g_{(\mu \nu)} \partial _{\delta} \partial _{\eta}
g_{(\rho \sigma)} \left\{ 2 g^{\gamma \delta} {X_{\alpha \beta}}
^{\rho \sigma} \frac{\partial H^{\eta}}{\partial g_{\mu \nu}} +
g^{\eta \delta} {X_{\alpha \beta}} ^{\mu \nu} \frac{\partial
H^{\gamma}}{\partial g_{\rho \sigma}} \right. \nonumber \\
& & \quad \mbox{}
+ (\delta _{\alpha} ^{\gamma} \delta _{\beta} ^{\delta} +
\delta _{\alpha} ^{\delta} \delta _{\beta} ^{\gamma}) G^{\rho
\sigma} \frac{ \partial H^{\eta}}{\partial g_{\mu \nu}} + \delta
_{\alpha} ^{\delta} \delta _{\beta} ^{\eta} G^{\mu \nu} \frac{
\partial H^{\gamma}}{\partial g_{\rho \sigma}} \nonumber \\
& & \quad \mbox{}
- \delta _{\beta} ^{\delta} {X_{\alpha}} ^{\gamma \rho
\sigma} \frac{\partial H^{\eta}}{\partial g_{\mu \nu}} - \delta
_{\beta} ^{\gamma} {X_{\alpha}} ^{\delta \rho \sigma}
\frac{\partial H^{\eta}}{\partial g_{\mu \nu}} - \delta _{\beta}
^{\eta} {X_{\alpha}} ^{\delta \mu \nu} \frac{\partial
H^{\gamma}}{\partial g_{\rho \sigma}} \nonumber \\
& & \quad \mbox{}
- \left. \delta _{\alpha} ^{\delta} {X_{\beta}} ^{\gamma
\sigma \rho} \frac{\partial H^{\eta}}{\partial g_{\mu \nu}} -
\delta _{\alpha} ^{\gamma} {X_{\beta}} ^{\delta \sigma \rho}
\frac{\partial H^{\eta}}{\partial g_{\mu \nu}} - \delta _{\alpha}
^{\eta} {X_{\beta}} ^{\delta \nu \mu} \frac{\partial
H^{\gamma}}{\partial g_{\rho \sigma}} \right\} ,
\label{dgddg2}
\end{eqnarraz}}%
where
\begin{equation}
G^{\rho \sigma} = \left\{
   \begin{array}{cl}
   g^{\rho \sigma} & \mbox{ if } \rho = \sigma , \\
   2 g^{\rho \sigma} & \mbox{ if } \rho \neq \sigma .
   \end{array} \right. \label{dgddg3}
\end{equation}
There are no $\partial _{\hat{\sigma}} \partial _{\hat{\sigma}}
g_{\rho \hat{\sigma}}$ terms in the Einstein equations, and no
first-degree $\partial _{\gamma} g_{\mu \nu}$ terms at all.
Derivatives like $\partial _{\gamma} g_{\mu \nu} \partial
_{\hat{\sigma}} \partial _{\hat{\sigma}} g_{\rho \hat{\sigma}}$,
for $\mu \leq \nu$ and $\rho \leq \hat{\sigma}$, are therefore
independent.  Setting the corresponding coefficients in
Eq.~(\ref{dgddg2}) to zero yields $\forall \alpha, \beta,
\gamma, (\mu \nu), (\rho \hat{\sigma})$
{\begin{eqnarraz}
0 &=& \left\{ 2 g^{\gamma  \hat{\sigma}} {X_{\alpha
\beta}}^{\rho
\hat{\sigma}} + (\delta _{\alpha} ^{\gamma} \delta _{\beta}
^{\hat{\sigma}} + \delta _{\alpha} ^{\hat {\sigma}} \delta
_{\beta} ^{\gamma}) G^{\rho  \hat{\sigma}} - \delta  _{\beta}
^{\hat{\sigma}} {X_{\alpha}}^{\gamma \rho \hat{\sigma}} - \delta
_{\beta} ^{\gamma} {X_{\alpha}}^{\hat{\sigma} \rho \hat{\sigma}}
\right. \nonumber \\
& & \quad \quad \left. \mbox{}
- \delta  _{\alpha} ^{\hat{\sigma}}
{X_{\beta}}^{\gamma \hat{\sigma} \rho} - \delta  _{\alpha}
^{\gamma} {X_{\beta}}^{\hat{\sigma} \hat{\sigma} \rho}
\right\} \frac{\partial H^{\hat{\sigma}}}{\partial g_{\mu \nu}}
\nonumber \\
& & \quad \mbox{}
+ \left\{ g^{\hat{\sigma} \hat{\sigma}} {X_{\alpha \beta}} ^{\mu
\nu} + \delta _{\alpha} ^{\hat{\sigma}} \delta _{\beta}
^{\hat{\sigma}} G^{\mu \nu} - \delta  _{\beta} ^{\hat{\sigma}}
{X_{\alpha}}^{\hat{\sigma} \mu \nu} - \delta  _{\alpha}
^{\hat{\sigma}} {X_{\beta}} ^{\hat{\sigma} \nu \mu} \right\}
\frac{\partial H^{\gamma}}{\partial g_{\rho \hat{\sigma}}} .
\label{dgddg4}
\end{eqnarraz}}%
We set $\alpha \neq \hat{\sigma}$, $\alpha \neq \gamma$, $\beta
\neq \hat{\sigma}$, and $\beta \neq \gamma$.  In three or more
dimensions, this yields $\forall \gamma, (\mu \nu), (\rho
\hat{\sigma})$
\begin{equation}
0 = g^{\hat{\sigma} \hat{\sigma}} {X_{\alpha \beta}} ^{\mu \nu}
\frac{\partial H ^{\gamma}}{\partial g_{\rho \hat{\sigma}}}
\label{dgddg5}.
\end{equation}
Since this must hold as an identity, we conclude that $\forall
\gamma, (\rho \hat{\sigma})$
\begin{equation}
0 = \frac{\partial H^{\gamma}}{\partial g_{\rho \hat{\sigma}}}
\label{d1}.
\end{equation}
That is, all partial derivatives of $H ^{\gamma}$ with respect to
components of the metric tensor vanish.

In two dimensions, the restrictions on indices before
Eq.~(\ref{dgddg5}) imply that $\alpha = \beta \neq \gamma =
\hat{\sigma}$.  From this we conclude that $\forall (\rho
\hat{\sigma})$
\begin{equation}
0 = \frac{\partial H^{\hat{\sigma}}}{\partial g_{\rho
\hat{\sigma}}} \label{dgddg6}.
\end{equation}
Now set $\alpha = \beta \neq \rho = \hat{\sigma}$ in
Eq.~(\ref{dgddg4}).  We find $\forall \gamma, (\mu \nu),
\hat{\alpha} \neq \hat{\sigma}$
\begin{equation}
0 = g^{\hat{\sigma} \hat{\sigma}} {X_{\hat{\alpha} \hat{\alpha}}}
^{\mu \nu} \frac{\partial H ^{\gamma}}{\partial g_{\hat{\sigma}
\hat{\sigma}}} ,
\label{dgddg7}
\end{equation}
from which we conclude that $\forall \gamma, \hat{\sigma}$
\begin{equation}
0 = \frac{\partial H^{\gamma}}{\partial g_{\hat{\sigma}
\hat{\sigma}}} \label{dgddg8}.
\end{equation}
Eqs.~(\ref{dgddg6}) and (\ref{dgddg8}) cover all partial
derivatives except $\partial H^1 / \partial g_{12}$.  But that is
easily seen to vanish by setting $\gamma = \rho = 1$,
$\hat{\sigma} = 2$, and $\alpha = \beta = \mu = \nu = 1$ in
(\ref{dgddg4}).  Therefore, Eq.~(\ref{d1}) also holds in two
dimensions.

From (\ref{dgddg1}), one easily sees that (\ref{d1}) is
sufficient for all $\partial g \partial \partial g$ terms to
vanish.
\subsection*{$\partial g \partial g \partial g$ terms}
$\partial g \partial g \partial g$ terms come from the
substitution of Eqs.~(\ref{phi3x}) and (\ref{phi4x}) in
(\ref{prvx}).  Since they are always multiplied by first or
second derivatives of $H^{\eta}$ with respect to $g_{\mu \nu}$,
they vanish identically due to (\ref{d1}).
\subsection*{$\partial \partial g$ terms}
There are explicit $\partial \partial g$ terms in
Eq.~(\ref{prvx}), and implicit ones through
Eq.~(\ref{phi4x}).
Regrouping all those terms and rearranging indices, we find that
they are given by
{\begin{eqnarraz}
\partial \partial g & \rightarrow & \frac{1}{2} \partial
_{\delta}
\partial _{\eta} g_{\rho \sigma} \left\{ \delta _{\alpha}
^{\rho} \delta _{\beta} ^{\sigma} \Phi  ^{\delta \eta} + \delta
_{\alpha} ^{\delta} \delta _{\beta} ^{\eta} \Phi ^{\rho \sigma} -
\delta _{\beta} ^{\eta} \delta _{\alpha } ^{\sigma} \Phi  ^{\rho
\delta } - \delta _{\alpha} ^{\eta} \delta _{\beta} ^{\sigma}
\Phi ^{\delta \rho} \mbox{\rule[-1.5ex]{0mm}{5ex}}
\right. \nonumber \\
& & \quad \mbox{}
+ 2 \delta _{\alpha} ^{\rho} \delta _{\beta} ^{\sigma}
g^{\gamma \delta} \partial _{\gamma} H^{\eta} - g^{\delta \eta}
\frac{\partial \Phi _{\alpha \beta}}{\partial g_{(\rho \sigma)}}
\nonumber \\
& & \quad \mbox{}
+ \delta _{\beta} ^{\delta} g^{\rho \sigma} \partial
_{\alpha} H^{\eta} + \delta _{\alpha} ^{\delta} g^{\rho \sigma}
\partial _{\beta} H^{\eta} - \delta _{\alpha} ^{\delta}
\delta_{\beta} ^{\eta} g^{\mu \nu} \frac{\partial \Phi _{\mu \nu
}}{\partial g_{(\rho \sigma)}} \nonumber \\
& & \quad \mbox{}
- \delta _{\beta} ^{\delta} \delta _{\alpha } ^{\sigma}
g^{\rho \gamma} \partial _{\gamma} H^{\eta} - \delta _{\alpha}
^{\sigma} g^{\rho \delta} \partial _{\beta} H^{\eta} +
\delta_{\beta} ^{\eta} g^{\gamma \delta} \frac{\partial \Phi
_{\gamma \alpha }}{\partial g_{(\rho \sigma)}} \nonumber \\
& & \quad \mbox{} - \left.
\delta _{\alpha} ^{\delta} \delta _{\beta}
^{\sigma} g^{\rho \gamma} \delta _{\gamma} H^{\eta} - \delta
_{\beta} ^{\sigma} g^{\rho \delta} \partial _{\alpha} H^{\eta} +
\delta_{\alpha} ^{\eta} g^{\gamma \delta} \frac{\partial \Phi
_{\gamma \beta }}{\partial g_{(\rho \sigma)}} \right\} .
\label{ddg1}
\end{eqnarraz}}%
There are no $\partial _{\hat{\sigma}} \partial _{\eta}
g_{\hat{\sigma} \hat{\sigma}}$ terms in the Einstein equations.
The coefficients of these terms in (\ref{ddg1}) must
therefore vanish.  To extract these coefficients, we must first
symmetrize the expression in curly brackets in $\delta$ and
$\eta$ (since $\partial _{\delta} \partial _{\eta}  =  \partial
_{\eta} \partial _{\delta}$).  With some cancellations, we get
$\forall \alpha, \beta, \eta, \hat{\sigma}$
{\begin{eqnarraz}
0 &=& \delta _{\alpha} ^{\hat{\sigma}} \left\{ -
g^{\hat{\sigma}
\gamma} \delta _{\beta} ^{\eta} \partial _{\gamma}
H^{\hat{\sigma}} - g^{\hat{\sigma} \eta} \partial _{\beta}
H^{\hat{\sigma}} - \delta _{\beta} ^{\eta} g^{\mu \nu}
\frac{\partial \Phi _{\mu \nu}}{\partial g_{\hat{\sigma}
\hat{\sigma}}} + g^{\gamma \eta} \frac{\partial \Phi _{\gamma
\beta}}{\partial g_{\hat{\sigma} \hat{\sigma}}} \right\}
\nonumber \\
& & \quad \mbox{} + \delta _{\beta} ^{\hat{\sigma}} \left\{ -
g^{\hat{\sigma} \gamma} \delta _{\alpha} ^{\eta} \partial
_{\gamma} H^{\hat{\sigma}} - g^{\hat{\sigma} \eta} \partial
_{\alpha} H^{\hat{\sigma}} - \delta _{\alpha} ^{\eta} g^{\mu \nu}
\frac{\partial \Phi _{\mu \nu}}{\partial g_{\hat{\sigma}
\hat{\sigma}}} + g^{\gamma \eta} \frac{\partial \Phi _{\gamma
\alpha}}{\partial g_{\hat{\sigma} \hat{\sigma}}} \right\}
\nonumber \\
& & \quad \mbox{}
+ 2 \delta  _{\alpha} ^{\hat{\sigma}} \delta _{\beta}
^{\hat{\sigma}} g^{\gamma \eta} \partial _{\gamma}
H^{\hat{\sigma}} - 2 g^{\hat{\sigma} \eta} \frac{\partial \Phi
_{\alpha \beta}}{\partial g_{\hat{\sigma} \hat{\sigma}}}
\nonumber \\
& & \quad \mbox{} + \delta _{\alpha}  ^{\eta}
\left\{g^{\hat{\sigma}
\hat{\sigma}} \partial _{\beta}  h^{\hat{\sigma}} +  g^{\gamma
\hat{\sigma}} \frac{\partial \Phi  _{\gamma \beta}}{\partial
g_{\hat{\sigma} \hat{\sigma}}} \right\} + \delta _{\beta}
^{\eta} \left\{g^{\hat{\sigma} \hat{\sigma}} \partial _{\alpha}
h^{\hat{\sigma}} +  g^{\gamma \hat{\sigma}} \frac{\partial \Phi
_{\gamma \alpha}}{\partial g_{\hat{\sigma} \hat{\sigma}}}
\right\} . \label{ddg2}
\end{eqnarraz}}%
In Eq.~(\ref{ddg2}),  we set $\alpha \neq \hat{\sigma}  \neq
\beta$ and $\alpha \neq  \eta \neq  \beta$.  We obtain $\forall
\alpha \neq \hat{\sigma}  \neq \beta$
\begin{equation}
0 = \frac{\partial \Phi _{\alpha \beta}}{\partial g_{\hat{\sigma}
\hat{\sigma}}} . \label{d2}
\end{equation}
Next, we set $\alpha = \beta = \eta \neq  \hat{\sigma}$.  We get
$\forall \hat{\alpha} \neq  \hat{\sigma}$
\begin{equation}
0 = g^{\hat{\sigma} \hat{\sigma}} \partial _{\hat{\alpha}}
H^{\hat{\sigma}} + g^{\gamma \hat{\sigma}} \frac{\partial \Phi
_{\gamma \hat{\alpha}}}{\partial g_{\hat{\sigma} \hat{\sigma}}} -
g^{\hat{\sigma} \hat{\alpha}} \frac{\partial \Phi _{\hat{\alpha}
\hat{\alpha}}}{\partial g_{\hat{\sigma} \hat{\sigma}}} .
\label{ddg3}
\end{equation}
Owing to Eq.~(\ref{d2}), this implies that $\forall \alpha
\neq
\hat{\sigma}$
\begin{equation}
0 = \partial _{\alpha} H^{\hat{\sigma}} + \frac{\partial \Phi
_{\hat{\sigma} \alpha}}{\partial g_{\hat{\sigma} \hat{\sigma}}} .
\label{d3}
\end{equation}
Eqs.~(\ref{d2}) and (\ref{d3}) are sufficient for
(\ref{ddg2}) to vanish identically.  This can be shown by
considering in turn all remaining cases, namely (i) $\alpha \neq
\hat{\sigma} \neq \beta$, $\alpha = \eta \neq \beta$; (ii)
$\alpha \neq \hat{\sigma} \neq \beta$, $\alpha \neq \eta =
\beta$; (iii) $\alpha = \hat{\sigma} \neq \beta$, $\alpha \neq
\eta \neq \beta$; (iv) $\alpha \neq \hat{\sigma} = \beta$,
$\alpha \neq \eta \neq \beta$; (v) $\alpha = \hat{\sigma} = \beta
\neq \eta$; (vi) $\alpha \neq \hat{\sigma} = \beta$, $\alpha \neq
\eta = \beta$; (vii) $\alpha = \hat{\sigma} \neq \beta$, $\alpha
= \eta \neq \beta$; (viii) $\alpha \neq \hat{\sigma} = \beta$,
$\alpha = \eta \neq \beta$; (ix) $\alpha = \hat{\sigma} \neq
\beta$, $\alpha \neq \eta = \beta$; and (x) $\alpha =
\hat{\sigma} = \eta  = \beta$.

We now go back to Eq.~(\ref{ddg1}), and consider terms of
the form $\partial _{\hat{\sigma}} \partial _{\hat{\sigma}}
g_{\rho \hat{\sigma}}$, for $\rho  \neq \hat{\sigma}$.  There are
no such terms in the Einstein equations.  Therefore, their
coefficients must vanish.  We get $\forall \alpha, \beta, \rho
\neq \hat{\sigma}$
{\begin{eqnarraz}
0 &=& - \delta _{\alpha} ^{\hat{\sigma}} \delta _{\beta}
^{\hat{\sigma}} \left\{ 2 g^{\rho  \gamma} \partial _{\gamma}
H^{\hat{\sigma}} + g^{\mu \nu} \frac{\partial \Phi _{\mu
\nu}}{\partial g_{\rho \hat{\sigma}}} \right\} + \delta _{\alpha}
^{\hat{\sigma}} \left\{ \delta _{\beta} ^{\rho} g^{\gamma
\hat{\sigma}} \partial _{\gamma} H^{\hat{\sigma}} + g^{\rho
\hat{\sigma}} \partial _{\beta} H^{\hat{\sigma}} +g^{\gamma
\hat{\sigma}} \frac{\partial \Phi _{\gamma \beta}}{\partial
g_{\rho \hat{\sigma}}} \right\} \nonumber \\
& & \quad \mbox{}
+ \delta _{\beta}  ^{\hat{\sigma}} \left\{ \delta
_{\alpha} ^{\rho} g^{\gamma \hat{\sigma}} \partial _{\gamma}
H^{\hat{\sigma}} + g^{\rho \hat{\sigma}} \partial _{\alpha}
H^{\hat{\sigma}} +g^{\gamma \hat{\sigma}} \frac{\partial \Phi
_{\gamma \alpha}}{\partial g_{\rho \hat{\sigma}}} \right\}
\nonumber \\
& & \quad \mbox{}
- g^{\hat{\sigma} \hat{\sigma}} \left\{ \delta _{\alpha}
^{\rho} \partial _{\beta} H^{\hat{\sigma}} + \delta _{\beta}
^{\rho} \partial _{\alpha} H^{\hat{\sigma}} + \frac{\partial \Phi
_{\alpha \beta}}{\partial g_{\rho \hat{\sigma}}} \right\} .
\label{ddg4}
\end{eqnarraz}}%
Let us first consider the case where $\alpha \neq \hat{\sigma}
\neq \beta$.  We get (for $\alpha$, $\beta$, and $\rho$ all
different from $\hat{\sigma}$)
\begin{equation}
0 = - g^{\hat{\sigma} \hat{\sigma}} \left\{ \delta _{\alpha}
^{\rho} \partial _{\beta} H^{\hat{\sigma}} + \delta _{\beta}
^{\rho} \partial _{\alpha} H^{\hat{\sigma}} + \frac{\partial \Phi
_{\alpha \beta}}{\partial g_{\rho \hat{\sigma}}}
\right\} . \label{ddg5}
\end{equation}
Setting $\rho = \beta \neq \alpha$ in Eq.~(\ref{ddg5}), we
get
$\forall \alpha, \hat{\rho}, \sigma \neq$
\begin{equation}
0 = \partial _{\alpha} H^{\sigma} + \frac{\partial \Phi _{\alpha
\hat{\rho}}}{\partial g_{\hat{\rho} \sigma}} . \label{d4}
\end{equation}
The case where $\rho = \alpha \neq \beta$ gives a similar result.
Next, setting $\rho = \alpha = \beta$ yields $\forall \hat{\rho}
\neq \sigma$
\begin{equation}
0 = 2 \partial _{\hat{\rho}} H^{\sigma} + \frac{\partial \Phi
_{\hat{\rho} \hat{\rho}}}{\partial g_{\hat{\rho} \sigma}} .
\label{d5}
\end{equation}
Finally, we have $\forall \alpha, \beta, \rho, \sigma$ such that
$\alpha \neq \sigma \neq \beta$, $\alpha \neq \rho \neq \beta$,
and $\rho \neq  \sigma$
\begin{equation}
0 = \frac{\partial \Phi _{\alpha \beta}}{\partial g_{\rho
\sigma}} . \label{d6}
\end{equation}
Note that Eqs.~(\ref{d4}) and (\ref{d6}) have no meaning in
two dimensions.

Eqs.~(\ref{d4}), (\ref{d5}), and (\ref{d6}) are sufficient
for Eq.~(\ref{ddg4}) to vanish identically.  This can be
shown by
considering in turn all remaining cases, namely (i) $\alpha =
\hat{\sigma} \neq  \beta$; (ii) $\alpha \neq \hat{\sigma} =
\beta$; and (iii) $\alpha = \hat{\sigma} =  \beta$.

At this point, it is very useful to define a function
$\tilde{\Phi} _{\alpha \beta}$ so that
\begin{equation}
\tilde{\Phi} _{\alpha \beta} = \Phi _{\alpha \beta} + g_{\alpha
\gamma} \partial _{\beta} H^{\gamma} + g_{\gamma \beta} \partial
_{\alpha} H^{\gamma} . \label{phitilde}
\end{equation}
Clearly, $\tilde{\Phi} _{\alpha \beta} = \tilde{\Phi} _{\beta
\alpha}$.  Owing to (\ref{d1}), Eqs.~(\ref{d2}), (\ref{d3}),
(\ref{d4}), (\ref{d5}), and (\ref{d6}) imply
\begin{equation}
\frac{\partial \tilde{\Phi} _{\alpha \beta}}{\partial
g_{\hat{\sigma} \hat{\sigma}}} = 0 \mbox{ if } \alpha \neq
\hat{\sigma} \neq \beta ; \label{d2a}
\end{equation}
\begin{equation}
\frac{\partial \tilde{\Phi} _{\hat{\sigma} \beta}}{\partial
g_{\hat{\sigma} \hat{\sigma}}} = 0 \mbox{ if } \hat{\sigma} \neq
\beta ; \label{d3a}
\end{equation}
\begin{equation}
\frac{\partial \tilde{\Phi} _{\alpha \hat{\rho}}}{\partial
g_{\hat{\rho} \sigma}} = 0 \mbox{ if } \alpha , \hat{\rho},
\sigma \neq ; \label{d4a}
\end{equation}
\begin{equation}
\frac{\partial \tilde{\Phi} _{\hat{\rho} \hat{\rho}}}{\partial
g_{\hat{\rho} \sigma}} = 0 \mbox{ if } \hat{\rho} \neq \sigma ;
\label{d5a}
\end{equation}
\begin{equation}
\frac{\partial \tilde{\Phi} _{\alpha \beta}}{\partial g_{\rho
\sigma}} = 0 \mbox{ if }
     \left\{ \begin{array}{c}
         \alpha \neq \sigma \neq \beta , \\
         \alpha \neq \rho \neq \beta , \\
         \rho \neq \sigma .
     \end{array} \right.
\label{d6a}
\end{equation}
Eqs.~(\ref{d2a})--(\ref{d6a}) mean that $\tilde{\Phi}
_{\alpha
\beta}$ is a function of $g_{\alpha \beta}$ (same indices) and
$x^{\lambda}$ alone.

The conditions we have obtained so far are necessary and
sufficient for the coefficients of $\partial _{\hat{\sigma}}
\partial _{\eta} g_{\hat{\sigma} \hat{\sigma}}$ and $\partial
_{\hat{\sigma}} \partial _{\hat{\sigma}} g_{\rho \hat{\sigma}}$
terms to vanish.  They do not, however, make all coefficients of
$\partial \partial g$ terms in Eq.~(\ref{ddg1}) equal to
zero.
And this is as it should be since, owing to Eq.~(\ref{eve}), not
all second derivatives of the metric tensor are independent.
\subsection*{$\partial g$ terms}
There are many $\partial g$ terms in Eq.~(\ref{prvx}).  They
come from (\ref{phi3x}) and (\ref{phi4x}), and from the fact that
$\Gamma _{\tau \gamma \alpha}$ is related to $\partial _{\alpha}
g_{\tau \gamma}$ through (\ref{christoffel}).  Regrouping all
terms and making use of (\ref{christoffel1}), we get
{\begin{eqnarraz}
\partial g & \rightarrow & \frac{1}{2} g^{\gamma \delta} g^{\tau
\rho} \left\{ \left[ \partial _{\alpha} \Phi _{\tau \gamma}
+ \partial _{\gamma} \Phi _{\tau \alpha} - \partial _{\tau} \Phi
_{\gamma \alpha} \right] \Gamma _{\rho \delta \beta}
 + \left[ \partial _{\beta} \Phi _{\rho \delta} + \partial
_{\delta} \Phi _{\rho \beta} - \partial _{\rho} \Phi _{\delta
\beta} \right] \Gamma _{\tau \gamma \alpha} \right. \nonumber \\
& & \qquad \mbox{} - \left. \left[
\partial _{\delta} \Phi _{\tau \gamma} + \partial
_{\gamma} \Phi _{\tau \delta}  - \partial _{\tau} \Phi _{\gamma
\delta} \right] \Gamma _{\rho \alpha \beta}
 - \left[ \partial _{\beta} \Phi _{\rho \alpha} + \partial
_{\alpha} \Phi _{\rho \beta} - \partial _{\rho} \Phi _{\alpha
\beta} \right] \Gamma _{\tau \gamma \delta} \right\} \nonumber \\
& & \quad \mbox{}
+ \frac{1}{2}  g^{\gamma  \delta} \left\{ (\partial
_{\gamma} \partial _{\delta} H^{\eta}) (\Gamma _{\alpha \beta
\eta} + \Gamma _{\beta \alpha \eta}) + (\partial _{\alpha}
\partial _{\beta} H^{\eta})
(\Gamma _{\gamma \delta \eta} + \Gamma _{\delta \gamma \eta})
\mbox{\rule[-1.5ex]{0mm}{5ex}} \right. \nonumber \\
& & \qquad \mbox{}
- (\partial _{\delta} \partial _{\beta} H^{\eta}) (\Gamma
_{\gamma \alpha \eta} + \Gamma _{\alpha \gamma \eta}) - (\partial
_{\alpha} \partial _{\gamma} H^{\eta}) (\Gamma _{\beta \delta
\eta} + \Gamma _{\delta \beta \eta}) \nonumber \\
& & \qquad \mbox{}
- 2 \partial _{\gamma} \left( \frac{\partial \Phi
_{\alpha \beta}}{\partial g_{(\pi \sigma)}} \right) (\Gamma _{\pi
\sigma \delta} + \Gamma _{\sigma \pi \delta}) \nonumber \\
& & \qquad \mbox{}
- \partial _{\alpha} \left( \frac{\partial \Phi _{\gamma
\delta} }{\partial g_{(\pi \sigma)}} \right) (\Gamma _{\pi \sigma
\beta} + \Gamma _{\sigma \pi \beta}) - \partial _{\beta} \left(
\frac{\partial \Phi _{\gamma \delta} }{\partial g_{(\pi \sigma)}}
\right) (\Gamma _{\pi \sigma \alpha} + \Gamma _{\sigma \pi
\alpha}) \nonumber \\
& & \qquad \mbox{}
+ \partial _{\delta} \left( \frac{\partial \Phi _{\gamma
\alpha} }{\partial g_{(\pi \sigma)}} \right) (\Gamma _{\pi \sigma
\beta} + \Gamma _{\sigma \pi \beta}) + \partial _{\beta} \left(
\frac{\partial \Phi _{\gamma \alpha} }{\partial g_{(\pi \sigma)}}
\right) (\Gamma _{\pi \sigma \delta} + \Gamma _{\sigma \pi
\delta}) \nonumber \\
& & \qquad \mbox{} + \left. \partial _{\gamma} \left(
\frac{\partial \Phi
_{\delta \beta} }{\partial g_{(\pi \sigma)}} \right) (\Gamma
_{\pi \sigma \alpha} + \Gamma _{\sigma \pi \alpha}) + \partial
_{\alpha} \left( \frac{\partial \Phi _{\delta \beta} }{\partial
g_{(\pi \sigma)}} \right) (\Gamma _{\pi \sigma \gamma} + \Gamma
_{\sigma \pi \gamma}) \right\} . \label{dg1}
\end{eqnarraz}}%
Let us substitute $\Phi _{\alpha \beta}$, as given in
Eq.~(\ref{phitilde}), in Eq.~(\ref{dg1}).  Since
$\tilde{\Phi}
_{\alpha \beta}$ is a function of $g_{\alpha \beta}$ and
$x^{\lambda}$ only, we can write
{\begin{eqnarraz}
 \partial _{\gamma} \left( \frac{\partial \tilde{\Phi}
_{\hat{\alpha} \hat{\beta}}}{\partial g_{(\pi \sigma)}} \right)
(\Gamma _{\pi \sigma \delta} + \Gamma _{\sigma \pi \delta}) &=&
\partial _{\gamma} \left( \frac{\partial \tilde{\Phi}
_{\hat{\alpha} \hat{\beta}}}{\partial g_{\hat{\alpha}
\hat{\beta}}} {X_{\hat{\alpha} \hat{\beta}}} ^{\pi \sigma}
\right) (\Gamma _{(\pi \sigma) \delta} + \Gamma _{(\sigma \pi)
\delta}) \nonumber \\
&=& \partial _{\gamma} \left( \frac{\partial \tilde{\Phi}
_{\hat{\alpha} \hat{\beta}}}{\partial g_{\hat{\alpha}
\hat{\beta}}} \right) (\Gamma _{\hat{\alpha} \hat{\beta} \delta}
+ \Gamma _{\hat{\beta} \hat{\alpha} \delta}) . \label{dg2}
\end{eqnarraz}}%
After cancellations and rearrangements, (\ref{dg1}) becomes
{\begin{eqnarraz}
\partial g & \rightarrow & \frac{1}{2} g^{\gamma \delta} g^{\tau
\rho} \left\{ \left[ \partial _{\hat{\alpha}} \tilde{\Phi} _{\tau
\gamma} + \partial _{\gamma} \tilde{\Phi} _{\tau \hat{\alpha}} -
\partial _{\tau} \tilde{\Phi} _{\gamma \hat{\alpha}} \right]
\Gamma _{\rho \delta \hat{\beta}}  + \left[ \partial
_{\hat{\beta}} \tilde{\Phi} _{\rho \delta} + \partial
_{\delta} \tilde{\Phi} _{\rho \hat{\beta}} - \partial _{\rho}
\tilde{\Phi} _{\delta \hat{\beta}} \right] \Gamma _{\tau \gamma
\hat{\alpha}} \right. \nonumber \\
& & \qquad \mbox{} - \left. \left[ \partial _{\delta}
\tilde{\Phi} _{\tau \gamma} +
\partial _{\gamma} \tilde{\Phi} _{\tau \delta}  - \partial
_{\tau} \tilde{\Phi} _{\gamma
\delta} \right] \Gamma _{\rho \hat{\alpha} \hat{\beta}}
 - \left[ \partial _{\hat{\beta}} \tilde{\Phi} _{\rho
\hat{\alpha}} + \partial
_{\hat{\alpha}} \tilde{\Phi} _{\rho \hat{\beta}} - \partial
_{\rho} \tilde{\Phi} _{\hat{\alpha}
\hat{\beta}} \right] \Gamma _{\tau \gamma \delta} \right\}
\nonumber \\
& & \quad \mbox{}
+ \frac{1}{2} g^{\gamma  \delta} \left\{ - 2 \partial
_{\gamma} \left( \frac{\partial \tilde{\Phi} _{\hat{\alpha}
\hat{\beta}}}{\partial g_{\hat{\alpha} \hat{\beta}}} \right)
(\Gamma _{\hat{\alpha} \hat{\beta} \delta} + \Gamma _{\hat{\beta}
\hat{\alpha} \delta}) \right. \nonumber \\
& & \qquad \mbox{}
- \partial _{\hat{\alpha}} \left( \frac{\partial
\tilde{\Phi} _{\gamma \delta} }{\partial g_{\gamma \delta}}
\right) (\Gamma _{\gamma \delta \hat{\beta}} + \Gamma _{\delta
\gamma \hat{\beta}}) - \partial _{\hat{\beta}} \left(
\frac{\partial \tilde{\Phi} _{\gamma \delta} }{\partial g_{\gamma
\delta}} \right) (\Gamma _{\gamma \delta \hat{\alpha}} + \Gamma
_{\delta \gamma \hat{\alpha}}) \nonumber \\
& & \qquad \mbox{}
+ \partial _{\delta} \left( \frac{\partial \tilde{\Phi}
_{\gamma \hat{\alpha}} }{\partial g_{\gamma \hat{\alpha}}}
\right)
(\Gamma _{\gamma \hat{\alpha} \hat{\beta}} + \Gamma
_{\hat{\alpha} \gamma \hat{\beta}}) + \partial _{\hat{\beta}}
\left( \frac{\partial \tilde{\Phi} _{\gamma \hat{\alpha}}
}{\partial g_{\gamma \hat{\alpha}}} \right) (\Gamma _{\gamma
\hat{\alpha} \delta} + \Gamma _{\hat{\alpha} \gamma \delta})
\nonumber \\
& & \qquad  \mbox{} + \left.
\partial _{\gamma} \left( \frac{\partial
\tilde{\Phi} _{\delta \hat{\beta}} }{\partial g_{\delta
\hat{\beta}}} \right) (\Gamma _{\delta \hat{\beta} \hat{\alpha}}
+ \Gamma _{\hat{\beta} \delta \hat{\alpha}}) + \partial
_{\hat{\alpha}} \left( \frac{\partial \tilde{\Phi} _{\delta
\hat{\beta}} }{\partial g_{\delta \hat{\beta}}} \right) (\Gamma
_{\delta \hat{\beta} \gamma} + \Gamma _{\hat{\beta} \delta
\gamma}) \right\} . \label{dg3}
\end{eqnarraz}}%
There are no $\partial g$ terms in the Einstein equations.  Their
coefficients must therefore vanish.  Since the transformation
$\partial _{\alpha} g_{\tau \gamma} \rightarrow \Gamma _{\tau
\gamma \alpha}$ is nonsingular, the coefficients of
Christoffel symbols must vanish.  So we isolate $\Gamma _{\lambda
\mu \nu}$ in (\ref{dg3}), and set to zero its coefficient,
symmetrized in $\mu$ and $\nu$ (since $\Gamma _{\lambda \mu \nu}
= \Gamma _{\lambda \nu \mu}$).  The result is $\forall
\hat{\alpha}, \hat{\beta}, \hat{\lambda}, \hat{\mu}, \hat{\nu}$
{\begin{eqnarraz}
0 &=& \left( \delta _{\hat{\alpha}} ^{\hat{\mu}} \delta
_{\hat{\beta}} ^{\hat{\nu}} + \delta _{\hat{\alpha}} ^{\hat{\nu}}
\delta _{\hat{\beta}} ^{\hat{\mu}} \right) \left\{ - g^{\gamma
\delta} g^{\tau \hat{\lambda}} \left( \partial _{\delta}
\tilde{\Phi} _{\tau \gamma} + \partial _{\gamma} \tilde{\Phi}
_{\tau \delta} - \partial _{\tau} \tilde{\Phi} _{\gamma \delta}
\right) \mbox{\rule[-2.5ex]{0mm}{7ex}} \right. \nonumber \\
& & \quad \quad \left.
\mbox{} + g^{\hat{\lambda} \gamma} \partial _{\gamma}
\left( \frac{\partial \tilde{\Phi} _{\hat{\lambda}
\hat{\alpha}}}{\partial g _{\hat{\lambda} \hat{\alpha}}} \right)
+ g^{\hat{\lambda} \gamma} \partial _{\gamma} \left(
\frac{\partial \tilde{\Phi} _{\hat{\lambda}
\hat{\beta}}}{\partial g _{\hat{\lambda} \hat{\beta}}} \right)
\right\} \nonumber \\
& & \quad \mbox{}
+ \left( \delta _{\hat{\alpha}} ^{\hat{\lambda}} \delta
_{\hat{\beta}} ^{\hat{\mu}} + \delta _{\hat{\alpha}} ^{\hat{\mu}}
\delta _{\hat{\beta}} ^{\hat{\lambda}} \right) \left\{ - 2
g^{\gamma \hat{\nu}} \partial _{\gamma} \left( \frac{\partial
\tilde{\Phi} _{\hat{\alpha} \hat{\beta}}}{\partial g
_{\hat{\alpha} \hat{\beta}}} \right) + g^{\gamma \hat{\nu}}
\partial _{\gamma} \left( \frac{\partial \tilde{\Phi} _{\hat{\nu}
\hat{\lambda}}}{\partial g _{\hat{\nu}  \hat{\lambda} }} \right)
\right\} \nonumber \\
& & \quad \mbox{}
+ \left( \delta _{\hat{\alpha}} ^{\hat{\lambda}} \delta
_{\hat{\beta}} ^{\hat{\nu}} + \delta _{\hat{\alpha}} ^{\hat{\nu}}
\delta _{\hat{\beta}} ^{\hat{\lambda}} \right) \left\{ - 2
g^{\gamma \hat{\mu}} \partial _{\gamma} \left( \frac{\partial
\tilde{\Phi} _{\hat{\alpha} \hat{\beta}}}{\partial g
_{\hat{\alpha} \hat{\beta}}} \right) + g^{\gamma \hat{\mu}}
\partial _{\gamma} \left( \frac{\partial \tilde{\Phi} _{\hat{\mu}
\hat{\lambda}}}{\partial g _{\hat{\mu}  \hat{\lambda} }} \right)
\right\} \nonumber \\
& & \quad \mbox{} + \delta  _{\hat{\alpha}} ^{\hat{\mu}} \left\{
g^{\hat{\nu} \delta} g^{\hat{\lambda} \rho} \left( \partial
_{\hat{\beta}} \tilde{\Phi} _{\rho \delta} + \partial _{\delta}
\tilde{\Phi} _{\rho \hat{\beta}} - \partial _{\rho} \tilde{\Phi}
_{\delta \hat{\beta}} \right) - 2 g^{\hat{\lambda} \hat{\nu}}
\partial _{\hat{\beta}} \left( \frac{\partial \tilde{\Phi}
_{\hat{\lambda} \hat{\nu}}}{\partial g _{\hat{\lambda}
\hat{\nu}}} \right) + g^{\hat{\lambda} \hat{\nu}} \partial
_{\hat{\beta}} \left( \frac{\partial \tilde{\Phi} _{\hat{\lambda}
\hat{\alpha}}}{\partial g _{\hat{\lambda} \hat{\alpha}}} \right)
\right\} \nonumber \\
& & \quad \mbox{} + \delta  _{\hat{\alpha}} ^{\hat{\nu}} \left\{
g^{\hat{\mu} \delta} g^{\hat{\lambda} \rho} \left( \partial
_{\hat{\beta}} \tilde{\Phi} _{\rho \delta} + \partial _{\delta}
\tilde{\Phi} _{\rho \hat{\beta}} - \partial _{\rho} \tilde{\Phi}
_{\delta \hat{\beta}} \right) - 2 g^{\hat{\lambda} \hat{\mu}}
\partial _{\hat{\beta}} \left( \frac{\partial \tilde{\Phi}
_{\hat{\lambda} \hat{\mu}}}{\partial g _{\hat{\lambda}
\hat{\mu}}} \right) + g^{\hat{\lambda} \hat{\mu}} \partial
_{\hat{\beta}} \left( \frac{\partial \tilde{\Phi} _{\hat{\lambda}
\hat{\alpha}}}{\partial g _{\hat{\lambda} \hat{\alpha}}} \right)
\right\} \nonumber \\
& & \quad \mbox{}
+ \delta  _{\hat{\beta}} ^{\hat{\mu}} \left\{ g^{\gamma
\hat{\nu} } g^{\tau \hat{\lambda} } \left( \partial
_{\hat{\alpha}} \tilde{\Phi} _{\tau \gamma} + \partial _{\gamma}
\tilde{\Phi} _{\tau \hat{\alpha}} - \partial _{\tau} \tilde{\Phi}
_{\gamma \hat{\alpha}} \right) - 2 g^{\hat{\lambda} \hat{\nu}}
\partial _{\hat{\alpha}} \left( \frac{\partial \tilde{\Phi}
_{\hat{\lambda} \hat{\nu}}}{\partial g _{\hat{\lambda}
\hat{\nu}}} \right) + g^{\hat{\lambda} \hat{\nu}} \partial
_{\hat{\alpha}} \left( \frac{\partial \tilde{\Phi}
_{\hat{\lambda} \hat{\beta}}}{\partial g _{\hat{\lambda}
\hat{\beta}}} \right) \right\} \nonumber \\
& & \quad \mbox{}
+ \delta  _{\hat{\beta}} ^{\hat{\nu}} \left\{ g^{\gamma
\hat{\mu} } g^{\tau \hat{\lambda} } \left( \partial
_{\hat{\alpha}} \tilde{\Phi} _{\tau \gamma} + \partial _{\gamma}
\tilde{\Phi} _{\tau \hat{\alpha}} - \partial _{\tau} \tilde{\Phi}
_{\gamma \hat{\alpha}} \right) - 2 g^{\hat{\lambda} \hat{\mu}}
\partial _{\hat{\alpha}} \left( \frac{\partial \tilde{\Phi}
_{\hat{\lambda} \hat{\mu}}}{\partial g _{\hat{\lambda}
\hat{\mu}}} \right) + g^{\hat{\lambda} \hat{\mu}} \partial
_{\hat{\alpha}} \left( \frac{\partial \tilde{\Phi}
_{\hat{\lambda} \hat{\beta}}}{\partial g _{\hat{\lambda}
\hat{\beta}}} \right) \right\} \nonumber \\
& & \quad
\mbox{} + \delta _{\hat{\alpha}} ^{\hat{\lambda}} g^{\hat{\mu}
\hat{\nu}} \left\{ \partial _{\hat{\beta}} \left( \frac{\partial
\tilde{\Phi} _{\hat{\mu} \hat{\alpha}}}{\partial g_{\hat{\mu}
\hat{\alpha}}} \right) + \partial _{\hat{\beta}} \left(
\frac{\partial \tilde{\Phi} _{\hat{\nu} \hat{\alpha}}}{\partial
g_{\hat{\nu} \hat{\alpha}}} \right) \right\} + \delta
_{\hat{\beta}} ^{\hat{\lambda}} g^{\hat{\mu}  \hat{\nu}} \left\{
\partial _{\hat{\alpha}} \left( \frac{\partial \tilde{\Phi}
_{\hat{\mu} \hat{\beta}}}{\partial g_{\hat{\mu} \hat{\beta}}}
\right) + \partial _{\hat{\alpha}} \left( \frac{\partial
\tilde{\Phi} _{\hat{\nu} \hat{\beta}}}{\partial g_{\hat{\nu}
\hat{\beta}}} \right) \right\} \nonumber \\
& & \quad \mbox{}
- 2 g^{\hat{\mu} \hat{\nu}} g^{\hat{\lambda} \rho} \left(
\partial _{\hat{\beta}} \tilde{\Phi} _{\rho \hat{\alpha}} +
\partial _{\hat{\alpha}} \tilde{\Phi} _{\rho \hat{\beta}} -
\partial _{\rho} \tilde{\Phi} _{\hat{\alpha} \hat{\beta}} \right)
. \label{dg4}
\end{eqnarraz}}%
In Eq.~(\ref{dg4}), we let $\hat{\alpha}  \neq \hat{\lambda}
\neq
\hat{\beta}$, $\hat{\alpha}  \neq \hat{\mu} \neq \hat{\beta}$,
and $\hat{\alpha} \neq \hat{\nu} \neq \hat{\beta}$.  We obtain,
$\forall \alpha, \beta$ and $\forall \lambda$ such that $\alpha
\neq \lambda \neq \beta$
\begin{equation}
0 = g^{\lambda \rho} \left\{\partial _{\beta} \tilde{\Phi} _{\rho
\alpha} + \partial _{\alpha} \tilde{\Phi} _{\rho \beta} -
\partial _{\rho} \tilde{\Phi} _{\alpha \beta} \right\} .
\label{d8}
\end{equation}

If the value of $\lambda$ was not restricted, we would easily
conclude that $\partial _{\beta} \tilde{\Phi} _{\kappa \alpha}$
vanished $\forall \beta, \kappa, \alpha$.  The conclusion
probably holds also with the restrictions on $\lambda$.  Indeed
it is unlikely that Eq.~(\ref{d8}) holds identically with
nonvanishing values of $\partial _{\beta} \tilde{\Phi} _{\kappa
\alpha}$, owing to the fact that $\tilde{\Phi} _{\kappa \alpha}$
does not involve metric components other than $g_{\kappa
\alpha}$.  If $\partial _{\beta} \tilde{\Phi} _{\kappa \alpha} =
0$, one immediately concludes that Eq.~(\ref{dg4}) holds
identically.

Thus, a calculation of $\tilde{\Phi} _{\kappa \alpha}$ could
proceed as follows: (i) attempt to prove that $\partial _{\beta}
\tilde{\Phi} _{\kappa \alpha} = 0$, with the result that
$\tilde{\Phi} _{\hat{\kappa} \hat{\alpha}} = \tilde{\Phi}
_{\hat{\kappa} \hat{\alpha}} (g_{\hat{\kappa} \hat{\alpha}})$;
(ii) find the conditions on $\tilde{\Phi}
_{\kappa \alpha}$ given by the vanishing of the independent
$\partial \partial g$ terms (not already discussed), of the
$\partial g \partial g$ terms and of the no-derivative terms
(including contributions coming from the elimination, by
Eq.~(\ref{eve}), of the dependent $\partial \partial g$
terms).
Such a calculation, especially step (ii), would be very
complicated.  Fortunately, there is a way around.
\section{Lie algebra of the generators}
The generator $v$ of a Lie symmetry of Einstein's equations has
the form of Eq.~(\ref{v}).  We have shown, in the last
section,
that $H^{\mu} = H^{\mu}(x^{\lambda})$ and that $\Phi_{\mu \nu}$
is given as in Eq.~(\ref{phitilde}) with $\tilde{\Phi}
_{\hat{\mu} \hat{\nu}} = \tilde{\Phi} _{\hat{\mu} \hat{\nu}}
(x^{\lambda}, g_{\hat{\mu} \hat{\nu}})$.  We can thus write
\begin{equation}
 v = H^{\mu}(x^{\lambda}) \frac{\partial}{\partial x^{\mu}} +
\left\{ - g_{\mu \gamma} \partial _{\nu} H^{\gamma}
(x^{\lambda}) - g_{\gamma \nu} \partial _{\mu} H^{\gamma}
(x^{\lambda}) + \tilde{\Phi} _{\mu \nu}(x^{\lambda}, g_{\mu \nu})
\right\} \frac{\partial}{\partial g_{(\mu \nu)}}. \label{va}
\end{equation}

Let $L$ be the set of all $v$ that generate Lie symmetries of
Einstein's equations.  Then $L$ is a Lie algebra.  It is well
known that general coordinate transformations are symmetries of
Einstein's equations.  Such transformations are generated by
\begin{equation}
v^{GCT} = f^{\alpha} \frac{\partial}{\partial x^{\alpha}} -
\left\{ g_{\alpha \gamma} \partial _{\beta} f^{\gamma} +
g_{\gamma \beta} \partial _{\alpha} f^{\gamma} \right\}
\frac{\partial}{ \partial g_{(\alpha \beta)}} ,
\label{vgct}
\end{equation}
where $f^{\alpha}$ is any function of $x^{\lambda}$.  Thus,
$v^{GCT}$ belongs to $L$.

Let $v$, given as in Eq.~(\ref{va}), belong to $L$.  Since
$L$ is a Lie algebra, $v - v^{GCT}$ also belongs to $L$.  Setting
$f^{\alpha} = H^{\alpha}$, we get that
\begin{equation}
\tilde{v} = \tilde{\Phi} _{\mu \nu}(x^{\lambda}, g_{\mu \nu})
\frac{\partial}{\partial g_{(\mu \nu)}} ,
\label{vtilde}
\end{equation}
with $\tilde{\Phi} _{\mu \nu}$ given by
Eq.~(\ref{phitilde}),
also belongs to $L$.  Moreover, the commutator of $\tilde{v}$
with any $v^{GCT}$ belongs to $L$.  This, we shall now show,
severely limits the form of the function $\tilde{\Phi} _{\mu
\nu}$.

It is not difficult to show that, for any $f^{\alpha}
(x^{\lambda})$ and any $\tilde{\Phi} _{\mu \nu} (x^{\lambda},
g_{\mu  \nu})$, the commutator of $\tilde{v}$ with $v^{GCT}$ is
given by
\begin{equation}
\left[ \tilde{v} , v^{GCT} \right] = F_{\mu \nu} \frac{\partial}{
\partial g _{(\mu \nu)}}, \label{comm}
\end{equation}
where for $\mu \leq  \nu$
\begin{equation}
 F_{\mu \nu} = - f^{\alpha} \partial _{\alpha} \tilde{\Phi}
_{\mu \nu} -  \tilde{\Phi} _{\mu \gamma} \partial _{\nu}
f^{\gamma} - \tilde{\Phi} _{\gamma \nu} \partial _{\mu}
f^{\gamma} + \left( g _{\alpha \gamma} \partial _{\beta}
f^{\gamma} + g _{\gamma \beta} \partial _{\alpha} h^{\gamma}
\right) \frac{\partial \tilde{\Phi} _{\mu  \nu}}{\partial
g_{(\alpha \beta)}} . \label{capf}
\end{equation}
Eq.~(\ref{capf}) will hold for all $\mu$ and $\nu$ if we set
$F_{\mu \nu} = F_{\nu \mu}$ when $\mu > \nu$.  Since
$\tilde{\Phi} _{\mu \nu}$ is a function of $x^{\lambda}$ and
$g_{\mu \nu}$ only, we have $\forall \hat{\mu}, \hat{\nu}$
\begin{equation}
 F_{\hat{\mu} \hat{\nu}} = - f^{\alpha} \partial _{\alpha}
\tilde{\Phi} _{\hat{\mu} \hat{\nu}} -  \tilde{\Phi} _{\hat{\mu}
\gamma} \partial _{\hat{\nu}} f^{\gamma} - \tilde{\Phi} _{\gamma
\hat{\nu}} \partial _{\hat{\mu}}  f^{\gamma} + \left( g _{\hat{
\mu} \gamma} \partial _{\hat{\nu}} f^{\gamma} + g _{\gamma
\hat{\nu}} \partial _{\hat{\mu}} h^{\gamma} \right)
\frac{\partial \tilde{\Phi} _{\hat{\mu} \hat{\nu}}}{\partial
g_{\hat{\mu} \hat{\nu}}} . \label{capfa}
\end{equation}

For the right-hand side of (\ref{comm}) to belong to $L$, it is
necessary that $F_{\hat{\mu} \hat{\nu}}$  be a function of
$x^{\lambda}$ and $g_{\hat{\mu} \hat{\nu}}$ only.  From
Eq.~(\ref{capfa}), this implies that
{\begin{eqnarraz}
 \left( - \tilde{\Phi} _{\hat{\mu} \gamma} + g_{\hat{\mu}
\gamma} \frac{\partial \tilde{\Phi} _{\hat{\mu}
\hat{\nu}}}{\partial g_{\hat{\mu} \hat{\nu}}} \right) \partial
_{\hat{\nu}} f^{\gamma} + \left( - \tilde{\Phi} _{\gamma
\hat{\nu}} + g_{\gamma \hat{\nu}} \frac{\partial \tilde{\Phi}
_{\hat{\mu} \hat{\nu}}}{\partial g_{\hat{\mu} \hat{\nu}}} \right)
\partial _{\hat{\mu}} f^{\gamma} &=& F_{\hat{\mu} \hat{\nu}}
(x^{\lambda}, g_{\hat{\mu} \hat{\nu}}) + f^{\alpha} \partial
_{\alpha} \tilde{\Phi} _{\hat{\mu} \hat{\nu}} \nonumber \\
& \equiv & \tilde{F} _{\hat{\mu} \hat{\nu}} (x^{\lambda},
g_{\hat{\mu} \hat{\nu}}) .
\label{ftilde}
\end{eqnarraz}}%
Since $f^{\gamma}$ is arbitrary, we can set $f^{\gamma}
(x^{\lambda}) = \delta _{\rho} ^{\gamma} f(x^{\sigma})$, for
$\rho$ and $\sigma$ fixed.  Letting $\dot{f}$ denote the
derivative of $f$ with respect to its argument, we obtain
$\forall \rho, \sigma, \hat{\mu},  \hat{\nu}$
\begin{equation}
\left( - \tilde{\Phi} _{\hat{\mu} \rho} + g_{\hat{\mu} \rho}
\frac{\partial \tilde{\Phi} _{\hat{\mu} \hat{\nu}}}{\partial
g_{\hat{\mu} \hat{\nu}}} \right) \delta _{\hat{\nu}} ^{\sigma}
\dot{f} + \left( - \tilde{\Phi} _{\rho \hat{\nu}} + g_{\rho
\hat{\nu}} \frac{\partial \tilde{\Phi} _{\hat{\mu}
\hat{\nu}}}{\partial g_{\hat{\mu} \hat{\nu}}} \right)
\delta_{\hat{\mu}} ^{\sigma} \dot{f} = \tilde{F} _{\hat{\mu}
\hat{\nu}} (x^{\lambda}, g_{\hat{\mu} \hat{\nu}}) . \label{fp}
\end{equation}

Let $\hat{\mu} = \hat{\nu} =  \sigma$.  We have $\forall
\hat{\mu}, \rho$
\begin{equation}
2 \left( - \tilde{\Phi} _{\hat{\mu} \rho} + g_{\hat{\mu} \rho}
\frac{\partial \tilde{\Phi} _{\hat{\mu}
\hat{\mu}}}{\partial g_{\hat{\mu} \hat{\mu}}} \right) \dot{f} =
\tilde{F} _{\hat{\mu} \hat{\mu}} (x^{\lambda}, g_{\hat{\mu}
\hat{\mu}}) . \label{fq}
\end{equation}
Redefining $\tilde{F}$ as $\tilde{F}/ 2 \dot{f}$, we get
\begin{equation}
- \tilde{\Phi} _{\hat{\mu} \rho} + g_{\hat{\mu} \rho}
\frac{\partial \tilde{\Phi} _{\hat{\mu}
\hat{\mu}}}{\partial g_{\hat{\mu} \hat{\mu}}} = \tilde{F}
_{\hat{\mu} \hat{\mu}} (x^{\lambda}, g_{\hat{\mu} \hat{\mu}}) .
\label{fr}
\end{equation}
Eq.~(\ref{fr}) holds identically for $\rho = \hat{\mu}$.
For $\hat{\rho} \neq \hat{\mu}$, it implies that
\begin{equation}
\tilde{\Phi} _{\hat{\mu} \hat{\rho}} = A_{\hat{\mu} \hat{\rho}}
g_{\hat{\mu} \hat{\rho}} + B_{\hat{\mu} \hat{\rho}} ,
\label{fs}
\end{equation}
where $A_{\hat{\mu} \hat{\rho}}$ and $B_{\hat{\mu} \hat{\rho}}$
are symmetric in $\hat{\mu}$ and $\hat{\rho}$ and are arbitrary
functions of $x^{\lambda}$.  Substituting (\ref{fs}) in
(\ref{fr}), we find that $\forall \hat{\rho} \neq \hat{\mu}$
\begin{equation}
\frac{\partial \tilde{\Phi} _{\hat{\mu} \hat{\mu}}}{\partial
g_{\hat{\mu} \hat{\mu}}} = A_{\hat{\mu} \hat{\rho}} ,
\label{ft}
\end{equation}
or
\begin{equation}
\tilde{\Phi} _{\hat{\mu} \hat{\mu}} = A_{\hat{\mu} \hat{\rho}}
g_{\hat{\mu} \hat{\mu}} + B_{\hat{\mu} \hat{\mu}} ,
\label{fu}
\end{equation}
where $B_{\hat{\mu} \hat{\mu}}$ is an arbitrary function of
$x^{\lambda}$.  Eqs.~(\ref{ft}) and (\ref{fu}) imply that,
for $\hat{\rho} \neq \hat{\mu}$, $A_{\hat{\mu} \hat{\rho}}$ is
independent of $\hat{\rho}$ and, since $A_{\hat{\mu} \hat{\rho}}$
is symmetric, independent of $\hat{\mu}$ also.  Defining $A =
A_{\hat{\mu} \hat{\rho}}$, we can write Eqs.~(\ref{fs}) and
(\ref{fu}) as
\begin{equation}
\tilde{\Phi} _{\mu \rho} = A(x^{\lambda}) g_{\mu \rho} + B_{\mu
\rho} (x^{\lambda}) ,
\label{fv}
\end{equation}
which now holds $\forall \rho, \mu$.  Eq.~(\ref{fv}) is a
necessary condition for the commutator of $\tilde{v}$  and
$v^{GCT}$ to belong to $L$.

Let us now substitute (\ref{fv}) in (\ref{d8}).
We obtain $\forall \alpha, \beta$ such that $\alpha
\neq \lambda \neq  \beta$
\begin{equation}
0 = - [\partial _{\rho} A] g^{\lambda \rho} g_{\alpha \beta} +
g^{\lambda \rho} \left\{ \partial _{\beta} B_{\rho \alpha} +
\partial _{\alpha} B_{\rho \beta} - \partial _{\rho} B_{\alpha
\beta}  \right\}  . \label{ga}
\end{equation}
This must hold as an identity.  Thus, terms made up of different
powers of the metric components must separately vanish.  That is,
\begin{equation}
0 = [\partial _{\rho} A] g^{\lambda \rho} g_{\alpha \beta}
\label{gb}
\end{equation}
and
\begin{equation}
0 = g^{\lambda \rho} \left\{ \partial _{\beta} B_{\rho \alpha} +
\partial _{\alpha} B_{\rho \beta} - \partial _{\rho} B_{\alpha
\beta}  \right\}  . \label{gc}
\end{equation}
From (\ref{gb}), and since the $g^{\lambda \rho}$ are independent
variables, we get $\forall \rho$
\begin{equation}
0 = \partial _{\rho} A \label{gd} .
\end{equation}
In three or more dimensions, Eq.~(\ref{gc}) implies
similarly that $\forall \alpha, \beta, \rho$
\begin{equation}
0 = \partial _{\beta} B_{\rho \alpha} + \partial _{\alpha}
B_{\rho \beta} - \partial _{\rho} B_{\alpha \beta}  . \label{ge}
\end{equation}
This, in turn, implies that $\forall \alpha, \beta, \rho$
\begin{equation}
0 = \partial _{\beta} B_{\rho \alpha} . \label{gf}
\end{equation}
Therefore, in three or more dimensions, $A$ and $B_{\mu \rho}$ in
Eq.~(\ref{fv}) must be constant.  That is,
\begin{equation}
\tilde{\Phi} _{\mu \rho} = A g_{\mu \rho} + B_{\mu \rho} .
\label{fvc}
\end{equation}

Now the Lie algebra property of $L$ once more limits the form of
$\tilde{\Phi} _{\mu \rho}$.  Starting from Eq.~(\ref{fvc}),
and going through an argument similar to the one between
Eqs.~(\ref{va}) and (\ref{fv}), we find that the commutator
(\ref{comm}) belongs to $L$ only if $\forall \mu, \rho$, $B_{\mu
\rho} = 0$.

In two dimensions, the situation is slightly more complicated.
Eq.~(\ref{gd}) still holds, but the constraint $\alpha \neq
\lambda \neq \beta$ before Eq.~(\ref{ga}) implies that
$\alpha = \beta$.  Eq.~(\ref{ge}) then reads, $\forall
\hat{\alpha}, \rho$
\begin{equation}
0 = 2 \partial _{\hat{\alpha}} B_{\rho \hat{\alpha}} - \partial
_{\rho} B_{\hat{\alpha} \hat{\alpha}}  . \label{gea}
\end{equation}
Let $\hat{\alpha} = \rho$.  We have $\partial _{\hat{\alpha}}
B_{\hat{\alpha}  \hat{\alpha}} = 0$, so that
\begin{equation}
B_{00} = B_{00} (x^1) , \qquad B_{11} =  B_{11}  (x^0).
\label{gh}
\end{equation}
With $\hat{\alpha} \neq \rho$, we have
\begin{equation}
2 \partial _0 B_{10} =  \partial _1 B_{00} , \qquad 2 \partial
_1 B_{01} =  \partial _0 B_{11} ,
\label{gi}
\end{equation}
whence, from (\ref{gh})
\begin{equation}
\partial _0 \partial _0 B_{10} = 0 = \partial _1  \partial _1
B_{10} . \label{gj}
\end{equation}
This implies that
\begin{equation}
B_{10} = a x^0 x^1 + b x^0 + c x^1 + d ,
\label{gk}
\end{equation}
where $a$, $b$, $c$, and $d$ are constants.
Substituting (\ref{gk}) in (\ref{gi}), we obtain
\begin{equation}
B_{00} = a (x^1)^2 + 2 b x^1 + f , \qquad B_{11} = a (x^0)^2 + 2
c x^0 + g , \label{gl}
\end{equation}
where $f$  and $g$ are constants.

The upshot is that, in two dimensions, Eq.~(\ref{fv}) holds
with $A$ constant and $B_{\mu \rho}$ given as in
Eqs.~(\ref{gk}) and
(\ref{gl}).  But again, the Lie algebra property of $L$ limits
the form of $\tilde{\Phi} _{\mu \rho}$, and we can show that
$\forall \mu, \rho$, $B_{\mu \rho} = 0$.

To sum up, we have shown that the generator of a Lie symmetry of
Einstein's vacuum equations in {\it N} dimensions necessarily has
the form of Eq.~(\ref{va}), with $H^{\mu} (x^{\lambda})$
arbitrary and $\tilde{\Phi} _{\mu \nu} = A g_{\mu
\nu}$.  The functions $H^{\mu} (x^{\lambda})$ correspond to
general coordinate transformations.  The constant $A$, on the
other hand, corresponds to uniform scale transformations of the
metric.  It is easy to check that such transformations leave
Einstein's equations invariant if and only if the cosmological
term vanishes. Provided the system (\ref{eve}) is
nondegenerate~\cite{olver}, we have thus shown that all Lie
symmetries of Einstein's vacuum equations in $N$ dimensions are
obtained from general coordinate transformations and, when the
cosmological term vanishes, uniform rescalings of the metric.
\end{document}